\documentclass[twocolumn,showpacs,preprintnumbers,amsmath,amssymb]{revtex4}

\usepackage{graphicx}
\usepackage{dcolumn}
\usepackage{bm}

\begin{document}

\preprint{APS/123-QED}

\title{Electron rest mass and energy levels of atoms in photonic crystal medium}

\author{Renat Kh. Gainutdinov}
 \email{Renat.Gainutdinov@ksu.ru}
\altaffiliation[Also at ]{Department of Physics, Kazan Federal
University, 18 Kremlevskaya St, Kazan 420008, Russia.}
\author{Marat A. Khamadeev}
\author{Myakzyum Kh. Salakhov}



\date{\today}



\begin{abstract}
Photonic crystals are periodic systems that consist of dielectrics
with different refractive indices. They are designed to act on
photons in contrast to semiconductor crystals whose periodicity
affects the motion of electrons. Here we consider the interaction
of an atomic electron with its own radiation field in the case
when the atom is placed in air voids of a photonic crystal and is
not in mechanical contact with the vibrational degrees of freedom
of the dielectric host. A strong modification of this interaction
from that in free space is shown to change the rest mass of the
electron, and this has a significant effect on the shift of the
atomic energy levels. This shift is investigated by using the
example of atomic hydrogen in a high-index-contrast photonic
crystal. The found effect may be of interest both from fundamental
and practical points of view.
\end{abstract}




\maketitle
\section{INTRODUCTION}
\label{sec:intro}  

Since the pioneering works of Yablonovitch \cite{yabl} and John
\cite{john87} photonic crystals (PCs) are a major field of
research. The variation of the photon density of states (DOS)
being a result of a modification of the electromagnetic fields in
PCs leads to quantum effects, including the coherent control of
the spontaneous emission \cite{quang97}, the appearance of
photon-atom bound states \cite{john90,john91,bay1,bay2}, the
non-Markovian character of radiative decay \cite{busch00},
enhanced quantum interference effects \cite{zhu97}, and the
localization of superradiance near the photonic band edge
\cite{john95}. It is important that the strong modification of the
DOS in PCs may provide new insight into the fundamental problems
of quantum electrodynamics (QED).

One of the most important effects of QED is the Lamb shift that
originally was defined as a splitting of $2P_{1/2}$ and $2S_{1/2}$
levels of atomic hydrogen. This motivated many efforts devoted to
the study of the Lamb shift in PCs
\cite{john90,john91,zhu00,Vats02,Li01,Wang04,Wang05}. It was found
that the interaction of an atom with its own radiation field can
be significantly modified in the PC medium, and as a result, can
lead to very large values of the Lamb shift, as compared to the
case of vacuum \cite{Wang04,Wang05}. The effect was investigated
by means of the standard methods that were successfully employed
for describing the Lamb shift in vacuum. However, one has to keep
in mind that because of the ultraviolet (UV) divergences, in the
theory of QED the Hamiltonian (Lagrangian) is only of formal
importance, and knowing them is not sufficient to compute results
for physical quantities. In addition, one needs to choose a
renormalization scheme which regulates the integrals and subtracts
the infinities. One of the key elements of the scheme is the
procedure of the mass renormalization that prescribes to subtract,
in solving the bound-state problem, the self-energy of a free
electron from that of the bound electron on the basis that the
electromagnetic mass of the electron must be included in its
observable mass. Here we show that in the case when we deal with
atoms in PCs this renormalization procedure removes not only
infinities appearing in the theory but also observable
contributions to physical quantities.

The origin of the problem is the fact that because of the
modification of the interaction of a charged particle in the PC
medium with its own radiation field, the electromagnetic mass of
an electron in the PC must differ from its electromagnetic mass in
vacuum, which is included into the observable mass of the
electron. This means that in the PC medium the rest mass of the
electron should change its value. We show that this change is
observable and gives rise to a significant shift of the energy
levels of an isolated atom in PCs provided it is not in mechanical
contact with the vibrational degrees of freedom of the dielectric
host. The effect is investigated by using the example of atomic
hydrogen in a high index-contrast photonic crystal.

\section{INFLUENCE OF AN ENVIRONMENT ON THE ELECTROMAGNETIC MASS OF THE ELECTRON}
\begin{figure}[h]
\begin{center}
\begin{tabular}{c}
\includegraphics[width=\linewidth]{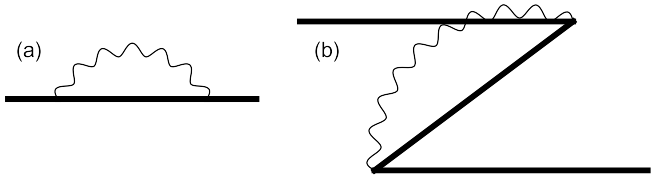}
\end{tabular}
\end{center}
\caption{The time-ordered diagrams describing the dominant
contribution to the Lamb shift. The thick line denotes the
electron (positron) propagating in the Coulomb field; the wavy
line denotes emission and reabsorption of a virtual photon.}
\label{sigma}
\end{figure}

The electromagnetic mass of the electron is its self-energy
associated with the interaction of the electron with its own
radiation field. Because of the UV divergences, this correction to
the electron mass is infinite. The problem is solved by using the
renormalization theory that implies that from the very beginning
the theory is formulated in terms of the physical charge and
electron mass including all radiative corrections, and
correspondingly the infinite electron self-energy contributions
are subtracted in computing physical quantities such as the Lamb
shift in atoms. This renormalization procedure proved to be very
successful in computing the QED corrections to the energy levels
of isolated atoms. However, as we show below, by using the example
of the atomic hydrogen in the case when atoms are removed from the
vacuum and placed in an environment in which the photonic density
of states is different from that of the vacuum, such a subtraction
leads to missing an observable correction to the electron rest
mass. The dominant contribution to the Lamb shift in hydrogenlike
atoms is given by the one-photon (one-loop) self-energy arising
from the processes in which a photon is emitted and then is
reabsorbed by a bound electron [these processes are described by
the diagram in Fig.~\ref{sigma}(a)] and from the processes in
which the electron in its final state first appears out of vacuum
together with a photon and a positron which then annihilate along
with the initial electron [these processes are described by the
diagram in Fig.~\ref{sigma}(b)].
In quantum electrodynamics the corresponding contribution to the
Lamb shift is given by the term that appears in the second-order
perturbation theory, and in the Furry picture can be written as
\begin{equation}
\Delta E_n  = \left\langle n \right|H_I \frac{1}{{E_n^{(0)}  -
H_0^F }}H_I \left| n \right\rangle, \label{de_o_loop}
\end{equation}
where $H_0^F$ is the unperturbed Dirac-Coulomb Hamiltonian in the
Furry picture ($H_0^F \left| n \right\rangle  = E_n^{(0)} \left| n
\right\rangle$), and $H_I  = \int {d^3 x} {\cal{H}}_I (t =
0,{\bf{x}})$, with ${\cal{H}}_I (t,{\bf{x}})$ being the
interaction Hamiltonian density
\[
{\cal{H}}_I (t,{\bf{x}}) = \frac{e}{2}A_\mu  (t,{\bf{x}})\left[
{\overline \Psi  (t,{\bf{x}}),\gamma ^\mu  \Psi (t,{\bf{x}})}
\right].
\]
It involves the quantized electromagnetic field $A_\mu (x)$ and
the quantized Dirac field $\Psi (x)$. The Furry picture is a kind
of interaction representation in which the unperturbed Hamiltonian
$H_0^F$ is the sum of the true free Hamiltonian $H_0$ and the
interaction Hamiltonian describing the interaction with an
external field, and provides the most convenient way to solve the
bound state problem in quantum electrodynamics.

Usually the self-energy correction to the energy levels of
hydrogenlike atoms is calculated by dividing the integral over
virtual photon energies, which is involved in (\ref{de_o_loop}),
into a low-energy range ($k<\Lambda$), within which one can treat
the electron nonrelativistically but must take into account
effects to all orders in the external field, and a high energy
range ($k>\Lambda$), within which the problem must be treated
relativistically but in the lowest order in the external field.
The parameter $\Lambda$ must be chosen to be much larger than the
atomic binding energies, but much less than typical electron
momenta (here and below we use the unit system where $\hbar  = c =
\varepsilon_0 = 1$)
\begin{equation}\label{<<l<<}
(Z\alpha )^2 m_e  \ll \Lambda  \ll (Z\alpha )m_e.
\end{equation}

Thus, the self-energy shift is the sum of two terms, the
high-energy (HE) term $\Delta E_n^ >$ and the low-energy (LE) term
$\Delta E_n^ <$. In the nonrelativistic approximation we may
neglect the contribution to the Lamb shift from the processes
described by the diagram depicted in Fig. \ref{sigma}(b), and as a
consequence, for $\Delta E_n^ <$ we get from Eq. (\ref{de_o_loop})
the following expression \cite{Bjork}:
\begin{equation}
{\Delta E_n^ <   = \frac{{2\pi\alpha }}{{3 m_e^2 }}\int_0^{\Lambda
} {\frac{{d^3 k}}{{2\left| {\bf{k}} \right|(2\pi )^3
}}\sum\limits_m {\frac{{\left| {\left\langle n
\right|{\bf{p}}\left| m \right\rangle } \right|^2 }}{{E_n  -
\left| {\bf{k}} \right| - E_m }}} } }, \label{de_<}
\end{equation}
where $\alpha$ is the fine-structure constant. The above
prescription of the renormalization theory tells us that the
self-energy of a free electron must be subtracted from the
self-energy of the bound electron given by Eq. (\ref{de_<}). The
contribution to the self-energy of the free electron comes from
the processes in which in time intervals between the emission and
reabsorption the electron is free and does not interact with the
Coulomb field. 
Since for a free electron only diagonal elements of the operator
${\bf p}^2$ differ from zero, the contribution to $\Delta E_n^ <$
given by Eq. (\ref{de_<}) from such processes is
\begin{equation}
\Delta E_{v,n}^<  = - \frac{\Delta m_{e}}{2 m_e^2} \left\langle n
\right|{\bf{p}}^2 \left| n \right\rangle, \label{de_vn_<}
\end{equation}
where
\begin{equation}
\Delta m_e = \frac{\alpha }{{p^2 \pi ^2 }}\sum\limits_{\lambda  =
1}^2 {\int\limits_0^\Lambda  {\frac{{d^3 k}}{{2\left| {\bf{k}}
\right|^2}}} } {{\left| {{\bf{p}} \cdot {\bf{\varepsilon
}}_\lambda  ({\bf{k}})} \right|^2 }}. \label{dm_free_<}
\end{equation}
As follows from Eq. (\ref{de_vn_<}), $\Delta m_e$ should be
regarded as a contribution to the electromagnetic mass of the
electron \cite{Schweber}. Thus, $\Delta m_e$ in Eq.
(\ref{de_vn_<}) is the electromagnetic mass of the electron and
therefore must be subtracted because the mass $m_e$ which we deal
with is the physical mass of the electron including all the
radiative corrections. This subtraction yields
\[
{\Delta E_n^ <   = \frac{\alpha }{{6\pi ^2 m_e^2}}\int_0^{\Lambda
} {\frac{{d^3 k}}{{2\left| {\bf{k}} \right|^2 }}\sum\limits_m
{\frac{{\left| {\left\langle n \right|{\bf{p}}\left| m
\right\rangle } \right|^2 }}{{E_n  - \left| {\bf{k}} \right| - E_m
}}} } } \left( {E_n  - E_m } \right).
\]
Adding the high-energy part $\Delta E_n^ >$, which is calculated
by using the corresponding Feynman diagrams of quantum
electrodynamics, to this term, we arrive \cite{Bjork} at the
ordinary expression for the one-loop self-energy Lamb shift
$\Delta E_n  = \Delta E_n^ <   + \Delta E_n^ >$ in a hydrogenlike
atom where the logarithmic dependence of $\Delta E_n^ <$ on
$\Lambda$ is compensated by that of $\Delta E_n^ >$.

In order to generalize the theory for describing the Lamb shift in
atomic hydrogen placed in PCs, one has to take into account the
influence of the PC medium on the propagation of virtual photons
that come into play in the process of the self-interaction of the
atomic electron. Correspondingly, in this case the wavy lines in
Fig.~\ref{sigma} should describe the propagation of photons in the
PC medium. Formally, carrying out the mass renormalization for the
electron in the PC medium should result in the subtraction of the
self-energy of the free electron modified by this medium from the
modified self-energy of the bound electron, and this subtraction
was used \cite{john90,john91,zhu00,Vats02,Li01,Wang04,Wang05} in
the studies of the Lamb shift in atoms placed in PCs. However,
this way of solving the problem leads to missing some important
contributions to energy levels from the self-interaction of atomic
electrons. In fact, the renormalization theory prescribes to also
add the subtracted term $\Delta m_e$ to the "bare" mass $m_0$ of
the electron in order to arrive at its physical mass
$m_e=m_0+\Delta m_e$. Here we mean one of the two approaches to
renormalization. It has the merit of a clear physical
interpretation, but the second approach, the method of
counterterms, is the one normally used in quantum field theory. In
the second approach the electron mass $m_e$ in the original
Lagrangian is regarded as the physical mass. To cancel the
contribution from the self-energy of the free electron an extra
term is added to the Lagrangian that is called the
mass-renormalization counterterm. The problem is that the value of
the electromagnetic mass of the electron in the PC medium should
differ from that in vacuum, and therefore the result of adding
this electromagnetic mass to its "bare" mass will not be the
physical mass. Obviously, the change of the value of the electron
mass $\delta m_{pc}$ is the difference between the values of the
electromagnetic masses in the PC medium $\Delta m_{pc}$ and the
electromagnetic mass $\Delta m_{e}$ in vacuum
\begin{equation}
\delta m_{pc}  = \Delta m_{pc}   - \Delta m_e.\label{dm_diff}
\end{equation}
Thus, the influence of the PC medium on the interaction of an
electron with its own radiation field results in the change in its
mass: $m_e \rightarrow m_{pc}=m_e+ \delta m_{pc}$. Actually the
above arguments are correct for any environment in which the
photonic density of states is different from that of the vacuum,
and allow one to conclude that the rest mass of the electron
placed in this environment should change its value.

\section{PHOTONIC CRYSTAL MEDIUM CORRECTIONS TO THE ELECTRON REST MASS}

Let us consider the problem of the change in rest mass of an
electron in the PC medium in more detail. Since the behavior of
photons in the PC medium differs from that in vacuum only in the
optical range of frequencies, from Eq. (\ref{dm_diff}) it follows
that only the low-energy part of the electron electromagnetic mass
is relevant for the problem. In this case the problem can be
solved nonrelativistically and it is convenient to choose the
Coulomb gauge that has the advantage that the radiation is
completely described by the vector potential $\mathbf{A}$. In this
gauge the nonrelativistic Hamiltonian for an electron in an
electromagnetic field may be written in the form
\begin{equation}\label{H_I_free}
H_{el}  = \frac{1}{{2m_e }}{\left[ {{\bf{p}} -
e{\bf{A}}({\bf{r}})} \right]}^2,
\end{equation}
where ${\bf{r}}$ is the position of the electron. The one-loop LE
contribution $\Delta m_e$ to the electromagnetic mass of an
electron in the Coulomb gauge takes the form
\begin{widetext}
\begin{equation}\label{dm_free_compl}
\Delta m_e^{}  =  - \frac{{2m_e^2 }}{{{\bf{p}}_{}^2
}}\sum\limits_{{\bf{p}}'} {\sum\limits_{{\bf{k}}\varepsilon
_\lambda  }^{} {\frac{{\left\langle {\bf{p}} \right|H_I \left|
{{\bf{p}}';{\bf{k}}{\bf{,}}\varepsilon _\lambda  } \right\rangle
\left\langle {{\bf{p}}';{\bf{k}}{\bf{,}}\varepsilon _\lambda  }
\right|H_I \left| {\bf{p}} \right\rangle }}{{\frac{{{\bf{p}}^2
}}{{2m_e }} - \frac{{{\bf{p}}'^2 }}{{2m_e }} - \left| {\bf{k}}
\right|}}} },
\end{equation}
\end{widetext}
where $H_I = -\frac{e}{{m_e }}{\bf{p}} \cdot {\bf{A}}$ and
$\left|{\bf{p}}; {{\bf{k}},\varepsilon _\lambda  } \right\rangle$
is a state with an electron with momentum ${\bf{p}}$ and a photon
with momentum ${\bf{k}}$ and polarization vector $\varepsilon
_\lambda$. Here and below, in order to deal with states of norm
$1$, we discretize the continuum by enclosing the particles in a
cubic box of volume $V$, and by imposing periodic boundary
conditions to obtain states having the same spatial dependence as
the states $\left| {{\bf{p}}} \right\rangle$. The final results
for physical quantities must not depend on $V$ provided it is
large enough. Since in our investigations we deal with the
electrodynamics within a PC it is natural to use its volume as the
normalization volume $V$.

Obviously, in describing the low-energy part of the electron
self-energy in the PC medium it is especially important to take
into account the Bloch structure of the photon states that arises
because of the periodicity of dielectric function
$\varepsilon(\textbf{r})$. This structure means that the photon
states can be expanded in a set of Bloch states $\left| {\bf{k}}n
\right\rangle$, which can be obtained by means of the plane-wave
expansion method \cite{sakoda}. By introducing the operators $\hat
a_{{\bf{k}}n}^ \dag$ and $\hat a_{{\bf{k}}n}$ that describe the
creation and annihilation of the photon in the state $\left|
{\bf{k}}n \right\rangle$ respectively ($\hat a_{{\bf{k}}n}^ \dag
\left| 0 \right\rangle=\left| {\bf{k}}n \right\rangle$ and  $\hat
a_{{\bf{k}}n} \left| {\bf{k}}n \right\rangle=\left| 0
\right\rangle$), we can construct a modified "free"\ Hamiltonian
$H^f_0= \sum_{\bf{k}n } \omega _{\bf{k}n} \hat a_{{\bf{k}}n}^\dag
\hat a_{{\bf{k}}n}$ and a quantized vector potential
\begin{widetext}
\begin{equation}
{\bf{A}}_{pc}({\bf{r}},t)=\sum_{{\bf{k}}n}[{\bf{\cal{A}}}_{{\bf{k}}n}({\bf{r}})
\hat a_{{\bf{k}}n} e^{-i \omega _{\bf{k}n} t } +
{\bf{\cal{A}}}^*_{{\bf{k}}n}({\bf{r}}) \hat a_{{\bf{k}}n}^\dag
e^{i \omega _{\bf{k}n}t }],\label{A_pc}
\end{equation}
\end{widetext}
where ${\bf{\cal{A}}}_{{\bf{k}}n}({\bf{r}}) = {\sqrt{1/V \omega
_{{\bf{k}}n}}}{\bf{E}}_{{\bf{k}}n} ({\bf{r}})$ with
${\bf{E}}_{{\bf{k}}n} ({\bf{r}})$ being the Bloch eigenfunctions
satisfying the following orthonormality condition
\begin{equation}\label{normal}
\int_V {d^3 r\varepsilon ({\bf{r}}){\bf{E}}_{{\bf{k}}n}
({\bf{r}}){\bf{E}}_{{\bf{k}}'n'}^* ({\bf{r}})}  = V \delta
_{{\bf{kk'}}} \delta _{nn'}.
\end{equation}
In this way we actually arrive at a modified Furry picture, in
which not only the interaction of an electron with the Coulomb
field but also the interaction of photons with the PC medium is
taken into account from the very beginning. Correspondingly, in
the interaction Hamiltonian (\ref{H_I_free}) the quantized vector
potential describing the free electromagnetic field in the PC
medium should be replaced with $A_{pc}(\bf{r},t)$ defined by Eq.
(\ref{A_pc}). With the vector potential defined in this way the
expression for the Hamiltonian (\ref{H_I_free}) is transformed to
\begin{equation}\label{H_I_pc}
H_{el}^{pc}  = \frac{1}{{2m_e }}\left[ {{\bf{p}} - e{\bf{A}}_{pc}
({\bf{r}})} \right]^2.
\end{equation}
Correspondingly, the expression for the LE part of the
electromagnetic mass of the electron in the PC medium takes the
form
\begin{widetext}
\begin{equation}\label{dm_pc_compl}
\Delta m_{pc}^{}  =  - \frac{{2m_e^2 }}{{{\bf{p}}_{}^2
}}\sum\limits_{{\bf{p}}'} {\sum\limits_{{\bf{k}}n}^{\omega
_{{\bf{k}}n}  < \Lambda } {\frac{{\left\langle {\bf{p}}
\right|H_I^{pc} \left| {{\bf{p}}';{\bf{k}},n} \right\rangle
\left\langle {{\bf{p}}';{\bf{k}},n} \right|H_I^{pc} \left|
{\bf{p}} \right\rangle }}{{\frac{{{\bf{p}}^2 }}{{2m_e }} -
\frac{{{\bf{p}}'^2 }}{{2m_e }} - \omega _{{\bf{k}}n} }}} } .
\end{equation}
Taking into account that for the above reason only the LE parts of
$\Delta m_e$ and $\Delta m_{pc}$ give a contribution to $\delta
m_{pc}$ defined by Eq. (\ref{dm_diff}), from this equation we get
\begin{equation}\label{dm_diff_compl}
\delta m_{pc}^{}  =  - \frac{{2m_e^2 }}{{{\bf{p}}_{}^2 }}\left(
{\sum\limits_{{\bf{p}}'} {\sum\limits_{{\bf{k}}n}^{\omega
_{{\bf{k}}n}  < \Lambda } {\frac{{\left\langle {\bf{p}}
\right|H_I^{pc} \left| {{\bf{p}}';{\bf{k}},n} \right\rangle
\left\langle {{\bf{p}}';{\bf{k}},n} \right|H_I^{pc} \left|
{\bf{p}} \right\rangle }}{{\frac{{{\bf{p}}^2 }}{{2m_e }} -
\frac{{{\bf{p}}'^2 }}{{2m_e }} - \omega _{{\bf{k}}n} }}} } }
\right. - \sum\limits_{{\bf{p}}'}
{\sum\limits_{{\bf{k}}\varepsilon _\lambda  }^{\left| {\bf{k}}
\right| < \Lambda } {\left. {\frac{{\left\langle {\bf{p}}
\right|H_I \left| {{\bf{p}}';{\bf{k}}{\bf{,}}\varepsilon _\lambda
} \right\rangle \left\langle
{{\bf{p}}';{\bf{k}}{\bf{,}}\varepsilon _\lambda  } \right|H_I
\left| {\bf{p}} \right\rangle }}{{\frac{{{\bf{p}}^2 }}{{2m_e }} -
\frac{{{\bf{p}}'^2 }}{{2m_e }} - \left| {\bf{k}} \right|}}}
\right)} }.
\end{equation}
\end{widetext}
Here the cutoff $\Lambda$ in expressions for $\Delta m_e$ and
$\Delta m_{pc}$ is removed because virtual HE photons emitted by
an electron in the PC medium propagate as if they were in vacuum.
For this reason the contributions to the first and second terms on
the right-hand part of Eq. (\ref{dm_diff_compl}) that come from
the self-interaction processes involving high-energy virtual
photons must compensate each other. The matrix element
$\left\langle {{\bf{p}}' ;{\bf{k}},n} \right|H_I^{pc} \left|
{{\bf{p}} } \right\rangle$ of the interaction Hamiltonian
$H_I^{pc}  = -\frac{e}{{m_e }}{\bf{p}} \cdot {\bf{A}}_{pc}$ can be
represented in the form
\begin{widetext}
\[
\left\langle {{\bf{p}}';{\bf{k}},n} \right|H_I^{pc} \left|
{{\bf{p}}} \right\rangle  = -\frac{e}{m_e}\int {d^3 r} \Psi
_{{\bf{p}}'}^* ({\bf{r}})[ - i\nabla _{\bf{r}}
{\bf{A}}_{{\bf{k}}n} ({\bf{r}})]\Psi _{\bf{p}}^{} ({\bf{r}}) =
\frac{e}{{m_e V^{3/2} \sqrt {\omega _{{\bf{k}}n} } }}\int {d^3 r}
e^{ - i{\bf{p}}'{\bf{r}}} [ i\nabla _{\bf{r}} {\bf{E}}_{{\bf{k}}n}
({\bf{r}})]e^{i{\bf{pr}}}
\]
\end{widetext}
with $\Psi _{{\bf{p}}} ({\bf{r}})$ being the normalized wave
function of the electron state $\Psi _{\bf{p}} ({\bf{r}}) =
\left\langle {{\bf{r}}} \mid {\bf{p}} \right\rangle$. Here we have
taken into account that $\Psi _{\bf{p}}= e^{i{\bf{p}}
{\bf{r}}}/\sqrt{V} $ for ${\bf{r}} \in V$ and $\Psi _{\bf{p}}= 0$
for ${\bf{r}} \notin V$. Taking also into account that
${\bf{E}}_{{\bf{k}}n} ({\bf{r}})$ can be expanded as
\[
{\bf{E}}_{{\bf{k}}n} ({\bf{r}}) = \sum\limits_{\bf{G}}
{{\bf{E}}_{{\bf{k}}n} ({\bf{G}})e^{i\left( {{\bf{k}} + {\bf{G}}}
\right) \cdot {\bf{r}}} }
\]
with ${\bf{G}}$ being the reciprocal lattice vector of the
photonic crystal (${\bf{G}} = N_1 {\bf{b}}_1  + N_2 {\bf{b}}_2 +
N_3 {\bf{b}}_3$, where ${\bf{b}}_i$ are primitive basis vectors of
a reciprocal lattice), for $\left\langle {{\bf{p}}' ;{\bf{k}},n}
\right|H_I^{pc} \left| {{\bf{p}} } \right\rangle$ we get
\[
\left\langle {{\bf{p}}';{\bf{k}},n} \right|H_I^{pc} \left|
{\bf{p}} \right\rangle  = -\frac{e}{m}\frac{1}{{\sqrt {V\omega
_{{\bf{k}}n} } }}\sum\limits_{\bf{G}} {{\bf{p}} \cdot
{\bf{E}}_{{\bf{k}}n} ({\bf{G}})} \delta _{{\bf{p}},{\bf{q}}},
\]
with ${\bf{q}} = {\bf{p}}' + {\bf{k}} + {\bf{G}}$. In the same
way, for $\left\langle {\bf{p}} \right|H_I^{pc} \left|
{{\bf{p}}';{\bf{k}},n} \right\rangle$ we find
\[
\left\langle {\bf{p}} \right|H_I^{pc} \left|
{{\bf{p}}';{\bf{k}},n} \right\rangle  =
-\frac{e}{m_e}\frac{1}{{\sqrt {V\omega _{{\bf{k}}n} }
}}\sum\limits_{\bf{G}} {{\bf{p}} \cdot {\bf{E}}^*_{{\bf{k}}n}
({\bf{G}})} \delta _{{\bf{p}},{\bf{q}}}.
\]
Correspondingly, for the matrix elements of the interaction
Hamiltonian in the free space, we have
\[
\left\langle {{\bf{p}}';{\bf{k}},n} \right|H_I^{} \left| {\bf{p}}
\right\rangle  = -\frac{e}{m_e}\frac{1}{{\sqrt {2V\left| {\bf{k}}
\right|} }}\sum\limits_\lambda  {{\bf{p}} \cdot \varepsilon
_\lambda  ({\bf{k}})} \delta _{{\bf{p}},{\bf{q}}},
\]
\[
\left\langle {\bf{p}} \right|H_I^{} \left| {{\bf{p}}';{\bf{k}},n}
\right\rangle  = - \frac{e}{m_e}\frac{1}{{\sqrt {2V\left| {\bf{k}}
\right|} }}\sum\limits_\lambda  {{\bf{p}} \cdot \varepsilon
_\lambda  ({\bf{k}})} \delta _{{\bf{p}},{\bf{q}}},
\]
with ${\bf{q}} = {\bf{p}}' + {\bf{k}}$. Substituting these matrix
elements of interaction Hamiltonians $H_I^{pc}$ and $H_I$ into Eq.
(\ref{dm_diff_compl}) yields
\begin{widetext}
\begin{equation}\label{dm_diff_compl_2}
\delta m_{pc}   =  - \frac{{2e^2 }}{{{\bf{p}}_{}^2 V}}\left(
{\sum\limits_{\bf{G}} {\sum\limits_{{\bf{k}}n}^{}
{\frac{1}{{\omega _{{\bf{k}}n} }}\frac{{\left| {{\bf{p}} \cdot
{\bf{E}}_{{\bf{k}}n} ({\bf{G}})} \right|^2 }}{{\frac{{{\bf{p}}^2
}}{{2m_e }} - \frac{{\left( {{\bf{p}} - {\bf{k}} - {\bf{G}}}
\right)^2 }}{{2m_e }} - \omega _{{\bf{k}}n} }}} } } \right. -
\sum\limits_{{\bf{k}}}\sum\limits_{\lambda  = 1}^2 {\left.
{\frac{1}{{2\left| {\bf{k}} \right|}}\frac{{\left| {{\bf{p}} \cdot
\varepsilon _\lambda ({\bf{k}})} \right|^2 }}{{\frac{{{\bf{p}}^2
}}{{2m_e }} - \frac{{\left( {{\bf{p}} - {\bf{k}}} \right)^2
}}{{2m_e }} - \left| {\bf{k}} \right|}}} \right)}.
\end{equation}
Electrons in air voids of a PC mainly are atomic electrons. In the
case of atomic hydrogen the momentum of the atomic electron is of
order $\alpha m_e$. In this case $-\omega$ is the predominant term
in the denominator of Eq. (\ref{dm_diff_compl_2}), and hence this
equation can be rewritten in the form
\begin{equation}\label{dm_diff_compl_25}
\delta m_{pc}   = \frac{{2e^2 }}{{{\bf{p}}_{}^2 V}}\left(
{\sum\limits_{\bf{G}} {\sum\limits_{{\bf{k}}n}^{} {\frac{{\left|
{{\bf{p}} \cdot {\bf{E}}_{{\bf{k}}n} ({\bf{G}})} \right|^2
}}{{\omega _{{\bf{k}}n}^2 }}} } } \right. -
\sum\limits_{{\bf{k}}}\sum\limits_{\lambda = 1}^2 {\left.
{\frac{{\left| {{\bf{p}} \cdot \varepsilon _\lambda ({\bf{k}})}
\right|^2 }}{{ 2{\bf{k}} ^2 }}} \right)}.
\end{equation}
Now in the expression of $\delta m_{pc}$ we can replace the
discrete sums by integrals $\sum\nolimits_{{\bf{k}}n} {}  \to
\frac{V}{{(2\pi )^3 }}\sum\nolimits_n {\int {d^3 k} }$ and
$\sum\nolimits_{{\bf{k}}} {}  \to \frac{V}{{(2\pi )^3 }}{\int {d^3
k} }$. In this way we get
\begin{equation}\label{dm_diff_compl_3}
\delta m_{pc}  = \frac{\alpha }{{\pi ^2 }}\left[ {\sum\limits_n^{}
{\int\limits_{FBZ} {\frac{{d^3 k}}{{\omega _{{\bf{k}}n}^2 }}} }
\sum\limits_{\bf{G}} {\left| {{\bf{\hat p}} \cdot
{\bf{E}}_{{\bf{k}}n} ({\bf{G}})} \right|^2 }  - \int\limits_{}
{\frac{{d^3 k}}{2{{\bf{k}}^2 }}} \sum\limits_{\lambda =
1}^2{{\left| {{\bf{\hat p}} \cdot \varepsilon _\lambda ({\bf{k}})}
\right|^2 }}} \right]
\end{equation}
with ${\bf{\hat p}} = {\bf{p}}/\left| {\bf{p}} \right|$. Thus, in
contrast to the Lamb shift, the PC medium correction to the
electron mass does not depend on the position of the electron in a
PC's air void. At the same time, this correction depends on the
direction unit vector ${\bf{\hat p}} = {\bf{p}}/\left| {\bf{p}}
\right|$ of the electron momentum. In order to describe the
"mean"\ correction, we may average ${\bf{\hat p}}$ over all solid
angles by assuming that this vector is randomly orientated in
space
\begin{equation}\label{dm_diff_compl_ave}
\delta m_{pc}^\Omega   \equiv \frac{1}{{4\pi }}\int {d\Omega
\delta m_{pc}^{} }  = \frac{\alpha }{{3\pi ^2 }}\left[
{\sum\limits_n^{} {\int\limits_{FBZ} {\frac{{d^3 k}}{{\omega
_{{\bf{k}}n}^2 }}} } {{\sum\limits_{\bf{G}} {\left|
{{\bf{E}}_{{\bf{k}}n} ({\bf{G}})} \right|^2 } }} - \int\limits_{}
{\frac{{d^3 k}}{{{\bf{k}}^2 }}} } \right].
\end{equation}
\end{widetext}

The fact that the PC medium correction to the electron mass does
not depend on the position of the electron in the PC's air void,
allows us to represent the correction in the form
\begin{equation}
\delta m_{pc}^\Omega  =  \frac{{4\alpha }}{{3\pi}}\int {d\omega
\frac{{{N}(\omega )-\omega ^2}}{{\omega ^2 }}},\label{dm_em3}
\end{equation}
where $N(\omega ) = N_{DOS} (\omega )D(\omega )$, $N_{DOS} (\omega
)$ is the photon density of states
\[
N_{DOS} (\omega ) = \frac{1}{{4\pi }}\sum\limits_n^{}
{\int\limits_{FBZ} {d^3 k} } \delta (\omega  - \omega _{{\bf{k}}n}
)
\]
and
\[
D(\omega ) = \sum\limits_{\bf{G}} {\left| {{\bf{E}}_{{\bf{k}}n}
({\bf{G}})} \right|^2 } _{\left| {\omega _{{\bf{k}}n}  = \omega }
\right.}.
\]
The DOS describes the number of states per interval of energy at
each energy level that are available to be occupied, and in vacuum
is equal to $\omega ^2$. In the limit $\varepsilon ({\bf{r}})
\rightarrow 1$ the right-hand part of Eq. (\ref{dm_em3}) must
vanish. It is easy to show that this is actually the case. In
fact, since
\begin{equation}
\sum\limits_{\bf{G}} {\left| {{\bf{E}}_{{\bf{k}}n} ({\bf{G}})}
\right|^2 }  = \frac{1}{V}\int\limits_V {d^3 r} \left|
{{\bf{E}}_{{\bf{k}}n} ({\bf{r}})} \right|^2,\label{norm_1}
\end{equation}
from Eq. (\ref{normal}) it follows that in the case when
$\varepsilon ({\bf{r}})$ approaches $1$, $\sum\limits_{\bf{G}}
{\left| {{\bf{E}}_{{\bf{k}}n} ({\bf{G}})} \right|^2 }$ approaches
$1$ as well. Because in this case $D(\omega )\approx 1$ and hence
$N (\omega ) \approx N_{DOS} (\omega )$, the right-hand part of
Eq. (\ref{dm_em3}) becomes equal to zero. Among other things this
serves as evidence of the correctness of the orthonormality
condition for ${{\bf{E}}_{{\bf{k}}n} ({\bf{r}})}$ that determine
the electromagnetic field in a PC via Eq. (\ref{A_pc})

Equation (\ref{dm_em3}) establishes the connection between the PC
medium correction to the electron mass and the DOS in a PC. For a
given PC the DOS as well as the function $D(\omega)$ can be
calculated numerically. However, in order to understand the
dependence of the shift of the rest mass of the electron on the
DOS in a PC, it is reasonable to use a model DOS which could
recapture the basic features of PCs. In Ref. \cite{Vats02}, for
example, a model DOS was proposed which recaptures the basic
quantitative features of a pseudogap, and has the form
\[
N_{DOS}(\omega ) = \omega ^2 \left[ {1 - h\exp \left( { -
\frac{{\left( {\omega  - \omega _0 } \right)^2 }}{{\sigma ^2 }}}
\right)} \right].
\]
However, as it is easy to see, the difference between the model
DOS and that in vacuum $\omega^2$ decreases exponentially as the
frequency $\omega$ increases or decreases from $\omega_0$. This
means that the behavior of photons in the medium of such a PC
differs from that in free space only in the frequency range around
$\omega_0$ with the width of the order of $\sigma$. In an actual
PC structure the DOS differs from that in free space in a much
wider frequency region in which the PC medium may be approximately
treated \cite{Li01} as an effective homogeneous medium with an
average dielectric constant $\bar \varepsilon  = \varepsilon \cdot
f + \left( {1 - f} \right)$ where $\varepsilon$ is the dielectric
constant of the host material and $f$ is the dielectric fraction
in the PC.

\begin{figure}
\begin{center}
\begin{tabular}{c}
\includegraphics[width=\linewidth]{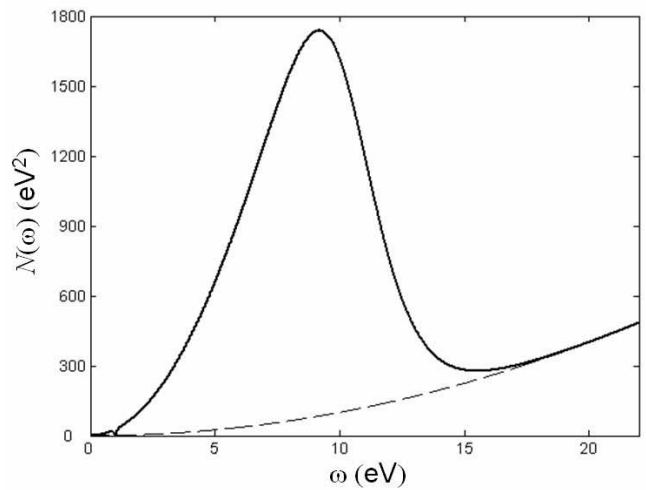}
\end{tabular}
\end{center}
\caption{The model $N(\omega)$ determined by Eq.
(\ref{DOS_vats_fermi}) with $n_{eff}=3$, $h=0.96$, $\sigma=0.07$
eV, $\mu=15$ eV, $\tau=0.01$ eV, and $\omega_0 = 1$ eV. The dashed
line denotes the free-space DOS.}\label{compound}
\end{figure}

These features of the PC medium are recaptured, for example, by
the following model function $N(\omega )$:
\begin{equation}
N(\omega ) = \omega ^2 n_{eff}^3 \left[ {1 - h\exp \left( { -
\frac{{\left( {\omega  - \omega _0 } \right)^2 }}{{\sigma ^2 }}}
\right)} \right] F(\omega ),\label{DOS_vats_fermi}
\end{equation}
where the factor $F(\omega ) = n_{eff}^{ - 3}  + (1 - n_{eff}^{ -
3} )/(\exp \left\{ {(\omega  - \mu )/\tau } \right\} + 1) $ with
$n_{eff}  \equiv \sqrt {\bar \varepsilon }$ allows one to take
into account that at high enough photon energies $N(\omega)$ must
approach the free-space DOS (Fig. \ref{compound}).

\begin{figure}[h]
\begin{minipage}[h]{1\linewidth}
\center{\includegraphics[width=1\linewidth]{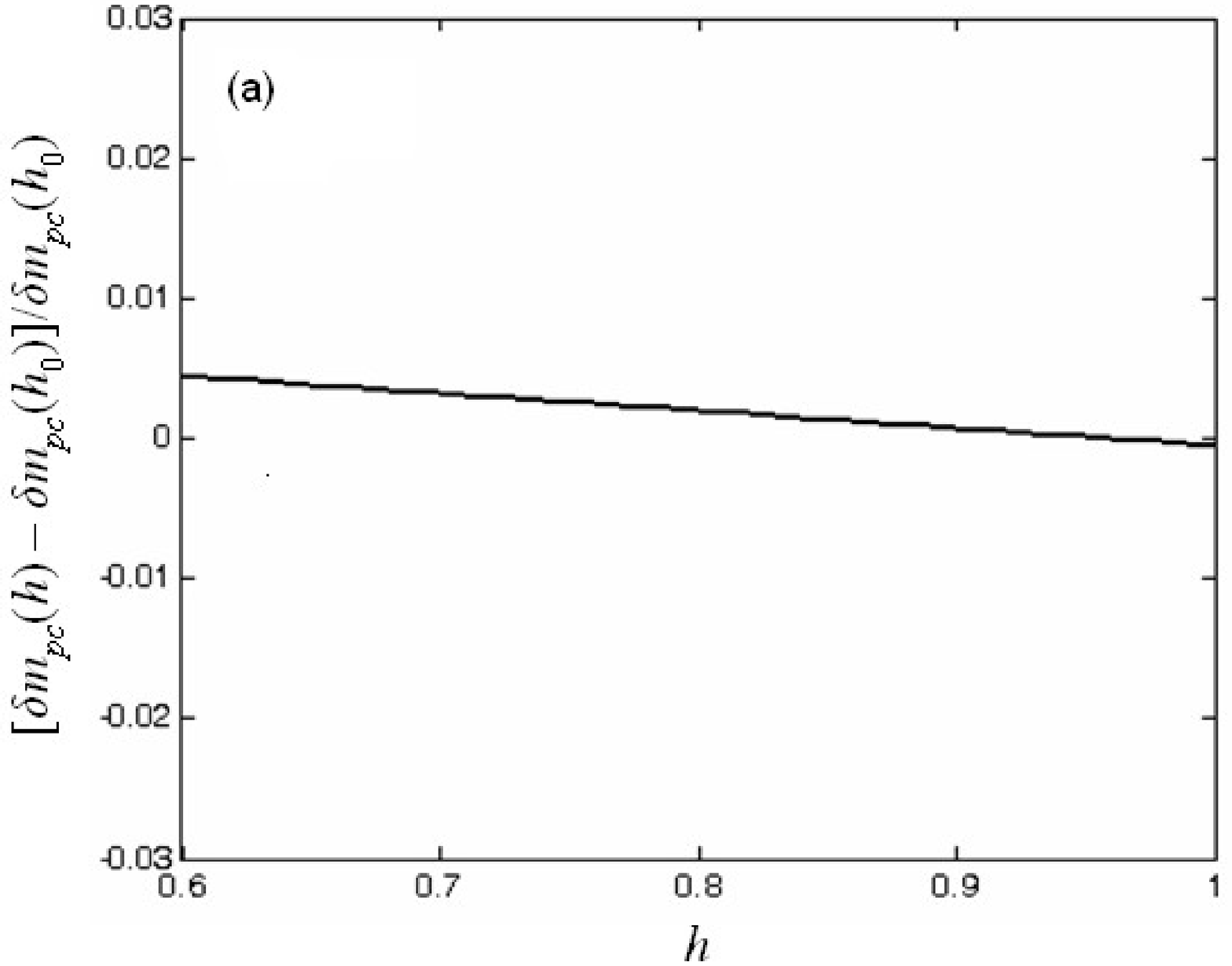}}
\end{minipage}
\hfill
\begin{minipage}[h]{1\linewidth}
\center{\includegraphics[width=1\linewidth]{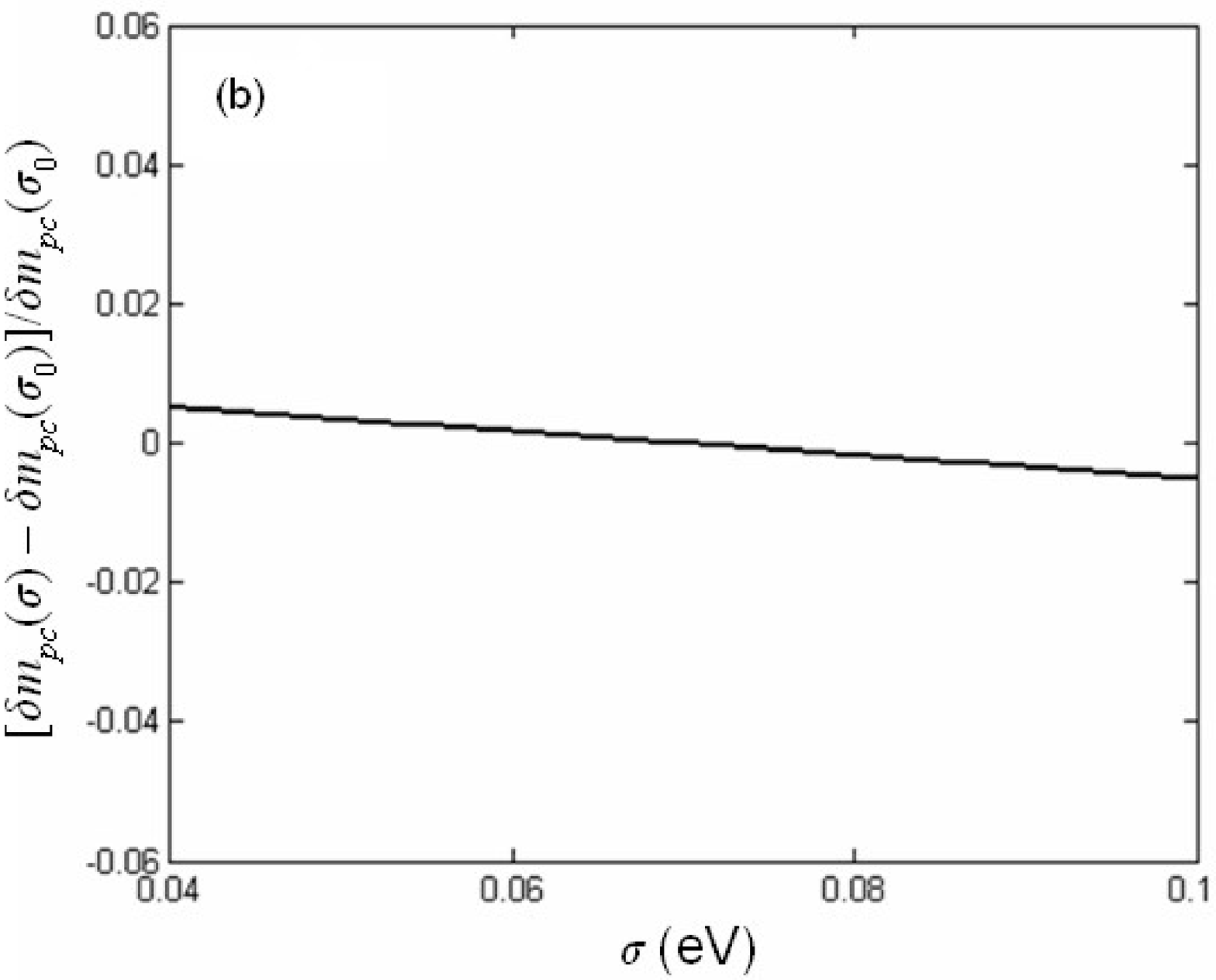}}
\end{minipage}
\caption{The dependence of the correction $\delta m_{pc}$ of the
electron rest mass in the PC medium on the values of the
parameters (a) $h$ and (b) $\sigma$ characterizing the pseudogap
for $n_{eff}=3$, $\mu = 15$ eV, $\tau=0.01$ eV, $\omega_0 = 1$ eV,
$h_0=0.96$, $\sigma_0=0.07$ eV, $\delta m_{pc}(h_0,\sigma_0)=2.4
\cdot 10^{-6} m_e$. Here $\delta m_{pc}$ is the mean value of the
electron mass correction defined by Eq. (\ref{dm_diff_compl_ave})
} \label{dm_fermi_ot}
\end{figure}

The results of our calculations of $\delta m_{pc}^\Omega$
displayed in Figs. \ref{dm_fermi_ot}(a) and \ref{dm_fermi_ot}(b)
show that the shift of the rest energy of the electron  in the PC
medium is insensitive to the values of the model parameters $h$
and $\sigma$ characterizing the pseudogap. This means that in the
one-loop approximation the contribution to the rest electron mass
that comes from the virtual photons with frequencies contained
within the pseudogap is negligible, and the effect depends mainly
on the behavior of the DOS at much higher frequencies. For given
values of the refractive index of the host dielectric and the
filling fraction $f$, this behavior is mainly determined by the
parameter $\mu$, whose value is chosen for the model DOS to
approach the free-space DOS at frequencies higher than the upper
bound $\omega_{op}$ of the optical frequency region. For the
parameters presented in the caption to Figs. \ref{dm_fermi_ot}(a)
and \ref{dm_fermi_ot}(b), our calculations have given $\delta
m_{pc}^\Omega = 2.4 \cdot 10^{-6} m_e$.

The change in the electron mass in a PC means that the energy of
the electron with the momentum ${\bf{p}}$ that in free space is
(in the nonrelativistic limit) $E_{\bf{p}} = m_e +
\frac{{\bf{p}}^2}{2 m_e}$ is changed to
\begin{equation}\label{E_p}
E_{\bf{p}} = (m_e + \delta m_{pc}) + \frac{{\bf{p}}^2}{2
(m_e+\delta m_{pc})}.
\end{equation}
As follows from Eq. (\ref{E_p}), the mass correction $\delta
m_{pc}$ depends on the orientation of the electron momentum in a
PC. Here it should be noted that we have derived the mass
correction from the contribution to the electron self-energy of
the form $ - \frac{{\delta m_{pc} }}{{m_e }}\frac{{{\bf{p}}^2
}}{{2m_e }}$, while there is the contribution to the self-energy
that does not depend on the electron momentum and directly
determines the correction $\delta m_{pc}$ to the rest energy. The
reason for this is that in describing the LE part of the
self-energy we used the nonrelativistic Hamiltonian
(\ref{H_I_pc}). The free part of this Hamiltonian is
$\frac{{{\bf{p}}^2 }}{{2m_e }}$, and the energy is understand as
the difference between the total energy and the rest energy
$\epsilon_{\bf{p}} = E_{\bf{p}} - m_e$. Thus, the self-interaction
correction to the electron energy obtained in this way does not
contain the correction to the rest energy, and for
$\epsilon_{\bf{p}}$ in the PC medium we have $\epsilon_{\bf{p}} =
\frac{{\bf{p}}^2}{2 (m_e+\delta m_{pc})}$. However, in order to
determine the total energy we must add the rest energy $m_e$
supplemented by the correction $\delta m_{pc}$. In this way we
arrive at Eq. (\ref{E_p}). It should be noted that the mass
dependence on orientation of the electron momentum in the PC is
not surprising. It is a consequence of anisotropy of the crystal.
In solid-state physics, crystal anisotropies result in the fact
that the effective mass of an electron depends on direction of the
electron momentum with respect to the crystal axes.

Obviously the change in the electron mass gives rise to the shift
of the energy levels of atoms. Let us consider the effect by using
the example of the atomic hydrogen. In the approximation where the
nucleus is assumed to be a point and infinitely massive the energy
levels of atomic hydrogen in the photonic crystal medium are given
by the solution of the Dirac equation for the energy eigenvalues.
For the energy of the atomic state $\left| {n,j,l,m} \right\rangle
$, we have $E_{nj}  = m_e R_{nj}$, with
\[
R_{nj}  = \left[ {1 + \left( {\frac{{\alpha }}{{n - \left( {j +
{\raise0.5ex\hbox{$\scriptstyle 1$} \kern-0.1em/\kern-0.15em
\lower0.25ex\hbox{$\scriptstyle 2$}}} \right) + \sqrt {\left( {j +
{\raise0.5ex\hbox{$\scriptstyle 1$} \kern-0.1em/\kern-0.15em
\lower0.25ex\hbox{$\scriptstyle 2$}}} \right)^2  - \alpha ^2 } }}}
\right)^2 } \right]^{ - 1/2}
\]
where the electron rest mass in vacuum is replaced by the mean
mass correction $\left\langle {\delta m_{pc} } \right\rangle$ that
is determined as
\begin{equation}\label{dm_mean}
\left\langle {\delta m_{pc} } \right\rangle  = \int {d^3 p} \Psi
_{njlm}^* ({\bf{p}})\delta m_{pc} ({\bf{\hat p}})\Psi _{njlm}^{}
({\bf{p}}).
\end{equation}
Thus, the change in the rest mass of the electron in the hydrogen
atom placed in a PC gives rise to the following shift of the
energy levels
\begin{equation}\nonumber
\delta E_{nj}^{pc}  = \left\langle \delta m_{pc} \right\rangle
R_{nj}.\label{E_1+}
\end{equation}
Here we do not take into account the corrections that appear
because of the modification of the Lamb shift caused by the change
in the electron rest mass. In other words, the shift $\delta
E_{nj}^{pc}$ of an energy level of atomic hydrogen in the PC
medium equals its free-space value multiplied by the ratio
$\left\langle \delta m_{pc} \right\rangle/m_e$. And in the
approximation, where the atomic nucleus is assumed to be infinity
massive, this is the case for any atom. This is because in this
approximation there is only one energy scale that is given by the
electron mass, and as a consequence, the energy of an atomic state
is the electron mass multiplied by some dimensionless factor.

The energy of the hydrogen state in free space $\left| a
\right\rangle  = \left| {n,j,l,m} \right\rangle$ may be written as
$E_a=m_e+\epsilon_{a}$ with $\epsilon_{a}= -
\frac{1}{2}\frac{\alpha^2 m_e }{n^2}+ {\rm O} (\alpha^4)$. Here
the rest energy part of $E_a$ is distinguished. The frequency
$\omega_{ab}$ of the transition between the state $\left| a
\right\rangle$ and the state $\left| b \right\rangle  = \left|
{n',j',l',m'} \right\rangle$ is given by
\begin{equation}\label{w_ab}
\omega_{ab}= \epsilon_{a} - \epsilon_{b}.
\end{equation}
The transition frequency $\omega_{ab}$ is equal to $\epsilon_{a} -
\epsilon_{b}$ because the rest energy contributions are the same
for both the states. The situation is dramatically changed in the
case when the atom is placed in the PC medium. As we have seen, in
this case the rest energy part of the total energy of the bound
electron depends on the orbital angular momentum and the angular
momentum $z$ component $m$. As a result, the rest energy parts of
the total energies of the states $\left| a \right\rangle$ and
$\left| b \right\rangle$ make a contribution to the transition
frequency
\begin{equation}\label{w_ab_pc}
\omega_{ab}^{pc}= {\left\langle {\delta m_{pc} } \right\rangle _a
- \left\langle {\delta m_{pc} } \right\rangle _b } +
\epsilon_{a}^{pc} - \epsilon_{b}^{pc}.
\end{equation}
Thus, in contrast to the free-space case, in the case of the PC
medium the rest energy parts of the total energies of the states
$\left| a \right\rangle$ and $\left| b \right\rangle$ make the
contribution to the frequency of the transition between these
states. Moreover, the difference between $\left\langle {\delta
m_{pc} } \right\rangle _a$ and $\left\langle {\delta m_{pc} }
\right\rangle _b$ makes a predominant contribution to the PC
correction $\delta \omega_{ab}^{pc}$ to the transition frequency
\[
\delta \omega _{ab}^{pc}  = \left( {\left\langle {\delta m_{pc} }
\right\rangle _a  - \left\langle {\delta m_{pc} } \right\rangle _b
} \right)\left( {1 + {\rm O}(\alpha ^4 )} \right)
\]
provided $l_a\neq l_b$ and/or $m_a\neq m_b$. Such a surprising
appearance of the contribution from the electron rest energy to
the atom transition frequencies in the case when the atom is
placed in the PC medium gives rise to the fact that the
corrections to these frequencies can be very significant. For
example, for all $S$ states, $\left\langle {\delta m_{pc} }
\right\rangle_{nS}$ coincide with $\delta m_{pc}^\Omega$, which is
determined by Eq. (\ref{dm_em3}), and in our model is found to be
$2.4 \times 10^{-6} m_e$. The value of the $\left\langle {\delta
m_{pc} } \right\rangle$ in $P$ states should be different but of
the same order of magnitude. Thus, the PC medium correction to the
frequencies of the transitions between the $S$ and $P$ states
should be of order $10^{-6} m_e$.

\section{OUTLOOK} \label{sec:conclusion}

We have shown that in the photonic crystal medium a quantum
electrodynamical effect of a new type takes place. The atoms
placed in a photonic crystal may be regarded as atoms in free
space, and as a result, they must have the ordinary line optical
spectrum. Photonic crystal medium affects only the self-radiation
field of these atoms. Unlike the free-space case in the photonic
crystal medium the interaction of an atomic electron with its own
radiation field that contributes to the mass manifests itself
explicitly and this gives rise to the change in its mass. We have
derived Eq. (\ref{dm_diff_compl_3}) that allows one to calculate
the mass correction $\delta m_{pc}$ for a given PC. From this
equation it follows that $\delta m_{pc}$ is independent of the
position of the electron in the PC voids but depends on the
direction of the electron momentum with respect to the photonic
crystal axes. This mass dependence on direction has a significant
effect on the structure of the atomic energy levels because it
give rise to the fact that the mean PC medium correction $\delta
m_{pc}$ in states with different orbital angular momenta and/or
angular momentum $z$ components are different. This in turn
results in the appearance of the term $\left\langle {\delta m_{pc}
} \right\rangle _a - \left\langle {\delta m_{pc} } \right\rangle
_b$ in the expression (\ref{w_ab_pc}) for the atomic transition
frequency $\omega_{ab}$. Thus, despite that the modification of
the interaction of the electron with its own radiation field in
the PC medium gives rise to relatively small corrections to the
rest electron mass (in our model they are of order $10^{-6} m_e$),
it results in the appearance of the term in the expressions for
atomic transition frequencies that is absent in the free-space
case. This terms makes the contribution to $\omega_{ab}$ of order
$10^{-6} m_e$, while the transition frequency in free space
determined by Eq. (\ref{w_ab}) is given by
$\omega_{ab}=\frac{\alpha^2}{2}m_e(\frac{1}{n_b^2}-\frac{1}{n_a^2})+
{\rm O}(\alpha ^4 ) $. Thus, the shifts of the energy levels of
the atom in the PC medium actually may be comparable to the atomic
transition frequency in free space, and for this the modification
of the electromagnetic field in this medium need not to be
extraordinary. This provides a way to drive the structure of the
atomic energy levels. In this way, in particular, light sources
with the line spectrum of a new type could be developed. Every
spectral line of such sources could be shifted in a wide range by
changing properties of a photonic crystal. The change in the rest
mass of an electron in the PC medium gives rise also to the change
of its magnetic properties. The magnetic moment $\mu$ of an
electron changes its value in the PC medium on the value $\delta
\mu_{pc} = -  \mu \delta m_{pc}/m_e$, where $\mu$ is the electron
magnetic moment in vacuum. This is important in view of the
observation of the effect under study as well as its applications.
It is also important that the mass correction $\delta m_{pc}$
carries valuable information about the electron self-interaction
that in solving other QED problem is hidden in the regularization
and renormalization procedure. This gives us the hope that
experimental investigations of the predicted effect might have
seen beyond the physics that is described by the QED
renormalization theory.


\bibliography{report}   
\bibliographystyle{spiebib}   

\acknowledgments     

This study was supported by Grant No. NSh-5289.2010.2 of the
President of the Russian Federation for Support of Leading
Scientific Schools and by the Federal Target Program "Research and
scientific-pedagogical cadres of innovative Russia", Grant No.
02.740.11.0428.

\end{document}